\documentclass[conference]{IEEEtran}
\IEEEoverridecommandlockouts
\usepackage[T1]{fontenc}
\usepackage{array}
\usepackage{booktabs}
\usepackage{cite}
\usepackage{amsmath,amssymb,amsfonts}
\usepackage{algorithmic}
\usepackage{graphicx}
\usepackage{textcomp}
\usepackage{xcolor}
\def\BibTeX{{\rm B\kern-.05em{\sc i\kern-.025em b}\kern-.08em
    T\kern-.1667em\lower.7ex\hbox{E}\kern-.125emX}}
\newcommand{\RomanNumeralCaps}[1]
    {\MakeUppercase{\romannumeral #1}}    
\begin{document}
% \title{Conference Paper Title*\\
\title{Integrated Fault Diagnosis and Control Design for DER Inverters using Machine Learning Methods}
\author{\IEEEauthorblockN{Forouzan Fallah}
\IEEEauthorblockA{\textit{Department of Electrical and }\\ \textit{Computer Engineering} \\
\textit{Tarbiat Modares University}\\
Tehran, Iran \\
forouzan.fallah@modares.ac.ir}
\and
\IEEEauthorblockN{Amin Ramezani}
\IEEEauthorblockA{\textit{Department of Electrical and }\\ \textit{Computer Engineering} \\
\textit{Tarbiat Modares University}\\
Tehran, Iran \\
ramezani@modares.ac.ir}
\and
\IEEEauthorblockN{Ali Mehrizi-Sani}
\IEEEauthorblockA{\textit{The Bradley Department of Electrical}\\\textit{ and Computer Engineering} \\
\textit{Virginia Tech}\\
Blacksburg, VA 24060 \\
mehrizi@vt.edu}
}
\maketitle
\begin{abstract}
This paper employs a supervised machine learning (ML) algorithm to propose an integrated fault detection and diagnosis (FDD) and fault-tolerant control (FTC) strategy to detect, diagnose, and classify the grid faults and correct the input voltage before affecting the grid-connected distributed energy resources (DER) inverters. This controller can mitigate the impact of grid faults on inverters by predicting and modifying the time series of their input voltage. Simulation results show the effectiveness of the proposed controller and evaluate its operating performance.
\end{abstract}
\section{Introduction}
\IEEEPARstart{T}{\lowercase{o}} meet the increasing demand for electrical energy, renewable energy sources, such as solar and wind, have become widely reliable alternatives to fossil fuels. With increased grid penetration of renewable energies, distributed generation (DG) power systems have drawn much attention because of their effectiveness and reliability [1].

DG systems should be able to inject high-quality current into the grid to satisfy the grid interconnection standard. Power electronic inverters are typically used in DG application systems because of the necessity of high efficiency and flexibility in both grid-tied and island modes of operation. Specifically, the critical aspect of utilizing a grid-connected inverter is to meet the power quality standard of DG, such as the IEEE 519 in the US [2].

Conventionally, for controlling grid-connected inverters, the linear controllers, such as the proportional-integral (PI) control, have been employed in the synchronous reference frame due to their simplicity and stability. However, the conventional PI decoupling controller does not give a proper inverter system with LCL filters. Even though this control scheme can ensure a moderate quality of injected current into the utility grid, the choice of controller gains often requires a process of trial and error [3].
\begin{figure}[h]
\centerline{\includegraphics[width=8cm,height=6cm, trim=4 4 4 4,clip]{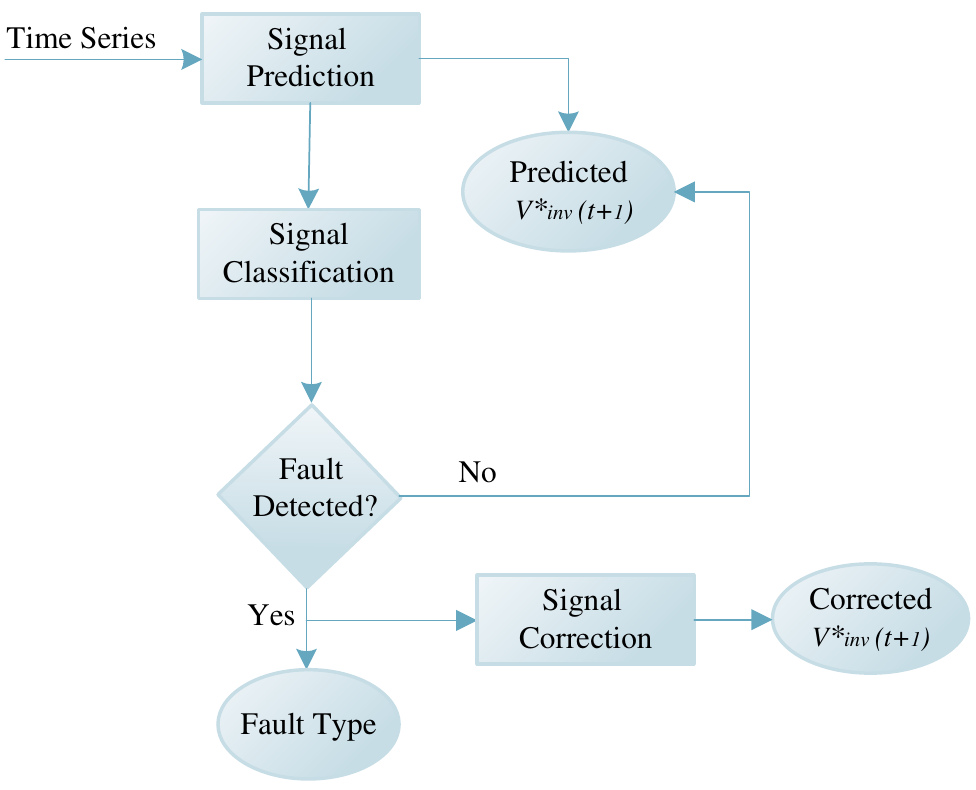}}
\caption{Flowchart of the proposed control algorithm.}
\vspace*{-3mm}
\label{fig}
\end{figure}
Additionally, Various types and levels of faults or anomalies, such as different types of grid faults, can affect DER inverters. in other words, DER Inverters are prone to suffering from instability issues that can lead to a further disaster. Some protection devices are commonly utilized to protect systems from catastrophic events, such as surge protection, anti-islanding protection, etc. [4]. However, early minor faults remain unprotected and are laborious and time-consuming to be detected manually, which may cause the reduction of the power generation efficiency and degradation of the system. Therefore, automatic fault detection and diagnosis (FDD) techniques are essential to detect these early faults for efficient, reliable, and safe operation. Moreover, fault-tolerant control (FTC) capability is the ability to continue service even under faulty conditions, which plays a predominant role in providing a high quality of power to the grid without any interruption.

Various FDD and FTC techniques have been put forward for power electronics converters in the last decade. These techniques are generally divided into model-based and data-driven approaches. The model-based methods are challenging to realize because the fault mathematical model is difficult to establish. Plus, an accurate and precise system model is hard to achieve. Compared with model-based methods, data-driven methods such as ML techniques are efficient and rely less on the circuit models, requiring only data. A neural network (NN) based fault diagnosis method for multilevel inverter drive was proposed in [5], in which the NN fault classifier was trained by the phase voltage after principal component analysis (PCA) dimension reduction, and the classification accuracy of the fault diagnosis is over 95\%. An intelligent anomaly identification technique for power electronics-based inverters is presented in [6], utilizing multi-class support vector machines (MSVM) for anomaly classification and localization. The proposed technique employs statistical features extracted from measurements for optimal learning of dual MSVM classifiers. An NN-based volt/VAR control strategy at the grid edge of a distribution system for inverter interfaced DERs is manifested in [7]. The NN implements optimal control according to approximate dynamic programming. Though these approaches have high performance, their control system design is not integrated and sometimes is regardless of the possibility of fault diagnosis or fault-tolerant capability. As a result, they are vulnerable to faults, or their fault-tolerant control strategy after detecting fault is the classical limited model-based approach.

In view of the advantages of ML techniques, In this paper, a new integrated ML-based FDD and FTC algorithm is proposed to detect, diagnose and classify all types of grid faults in a grid-connected DER inverter and correct the faulty voltage before affecting it and preventing system failure. Fig. 1 shows the flowchart of the proposed control algorithm.

The paper is organized as follows. Section II presents the DER inverter model. Section III reviews the proposed integrated ML-based FDD and FTC. Section IV presents simulation results. Finally, the paper concludes with a summary of the main points.
% \vspace*{-2.5mm}
\section{STUDY SYSTEM DESCRIPTION}
\subsection{General Problem Statement}
The principal objective of the proposed work is divided into FDD and FTC in the grid-tied three-phase DER inverter. The paper's novelty resides in integrating FDD and FTC using ML methods without the system model. The FDD process comprises of the following steps:

Step 1: Simulate 11 types of grid faults include single-phase, two-phase, two-phase to ground, three-phase, three-phase to ground.

Step 2: Measure and record inverter current, inverter input voltage and grid voltage for fault and normal states to create dataset. In this research, we have collected 800000 data samples for each training and test set.

Step 3: Use 70\% of training data to train the Long Short-Term Memory (LSTM) network and the K-nearest-neighbor (KNN) classifier to detect and identify faults, and other 30\% to validate the models. Further, 100\%  of entirely new data for the testing turn.

In continuation, the FTC process involves of the following steps:

Step 4: Use normal data set (no fault exists) to train and validate an NN-based multi-output regression to correct voltage signal in case of fault occurrence.

Step 5: Test the integrated controller with totally new data for different grid faults to verify the FDD and FTC algorithm.
\begin{figure}
\centerline{\includegraphics[width=9cm,height=4cm, trim=4 4 4 4,clip]{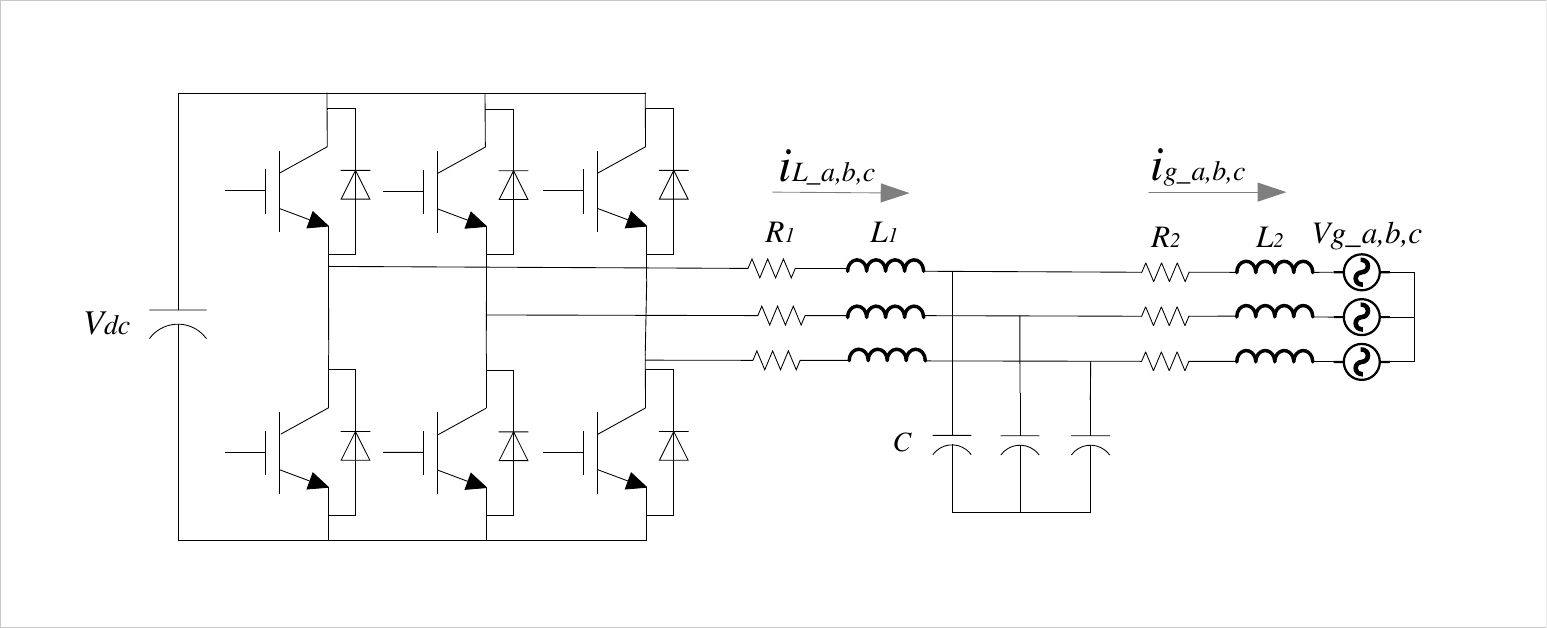}}
\caption{Configuration of a three-phase grid-connected inverter with an LCL filter.}
\vspace*{-3mm}
\label{fig}
\end{figure} 
\subsection{Inverter Model}
Fig. 2 shows a configuration of a three-phase grid-connected DER inverter with an LCL filter; in which $V_{dc}$ denotes the dc voltage across the DC-link capacitor that can be connected to a PV array via dc/dc converters or a wind generator via a dc/ac converter, a three-phase voltage source ${V_g}_{a,b,c}$ represents the three-phase ac voltage at the Point of Common Coupling (PCC). $R_1, R_2, L_1,$ and $L_2$ are the filter resistances and filter inductances, respectively, and C is the filter capacitance. ${V_{inv}}_{a,b,c}$ stands for the three-phase inverter output voltage in the ac system. ${i_L}_{a,b,c}$ represents the three-phase current flowing through the inductor, and ${i_g}_{a,b,c}$ signifies the three-phase current flowing into the grid at the PCC. 
\vspace*{-7mm}
\begin{figure}[h]
\centerline{\includegraphics[width=9cm,height=8cm, trim=4 4 4 4,clip]{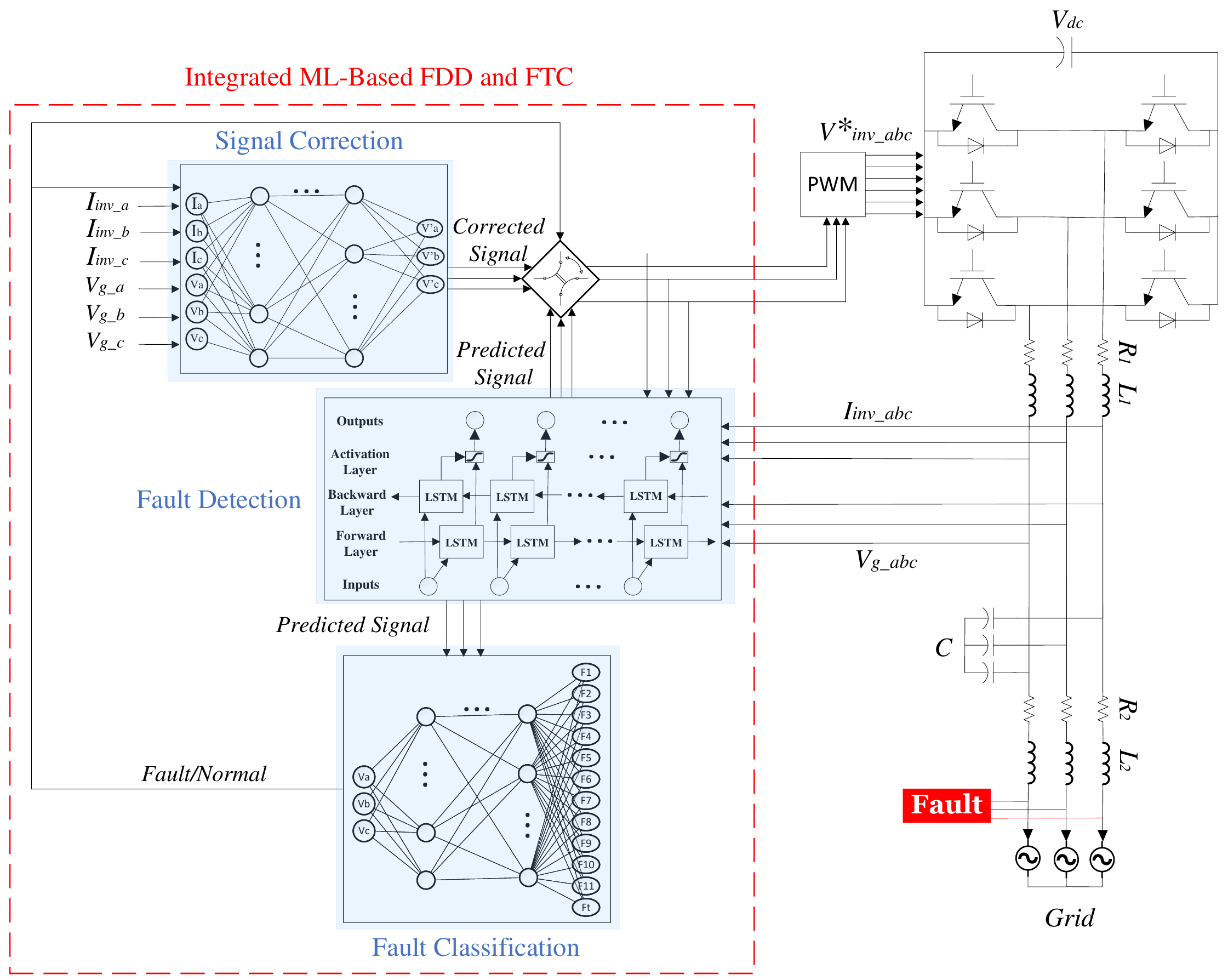}}
\caption{Structure of proposed FDD and FTC scheme.}
\vspace*{-1mm}
\label{fig}
\end{figure} 
\section{Integrated FDD and FTC algorithm}
In Fig. 3 the proposed control scheme is presented. This integrated controller consists of three parts. 
\begin{itemize}
    \item LSTM network
    \item KNN classifier
    \item NN-based multi-output regression 
\end{itemize}

The first part of the control scheme is responsible for predicting the next step of the PWM unit's input voltage which generates the inverter control signal. Then, in the second part, the predicted signal will be classified into 11 classes of fault or normal class. Finally, the third part of the control scheme will be activated and responsible for correcting the control signal in case of any fault.
% \begin{figure}[htbp]
% \centerline{\includegraphics[width=9cm,height=6cm]{training.pdf}}
% \caption{Example of a figure caption.}
% \label{fig}
% \end{figure}
\begin{figure}[h]
\centerline{\includegraphics[width=9cm,height=6cm, trim=4 4 4 4,clip]{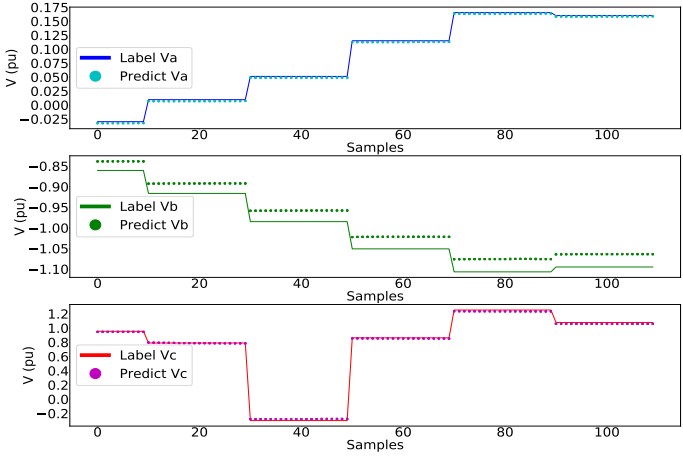}}
\caption{Signal prediction using LSTM network.}
\vspace*{-2mm}
\label{fig}
\end{figure}
\subsection{Fault Detection}
The I/O signals in inverters are time-series data sampled with sampling time $T$. The LSTM network has become the deep learning model of choice for sequential and temporal data because of its ability to learn long-range patterns. Relying on the proposed LSTM network for anomaly detection in [8], and utilizing the grid voltage ${V_g}_{a,b,c}(t)$, the inverter current ${I_{inv}}_{a,b,c}(t)$ and the inverter input voltage  ${{V^*}_{inv}}_{a,b,c}(t)$ as inputs, the LSTM network predicts next step of the inverter input voltage ${{V^*}_inv}_{a,b,c}(t+1)$. In fact, this part of the controller allows detecting the grid fault occurrence and makes it possible to prevent it before the inverter failure. Fig. 4 shows the prediction results on test data. It should be noted that, in the first step, ${{V^*}_{inv}}_{a,b,c}(t)$ has not been generated yet; hence, it can be set by selecting a random value in the range $[-1, 1]$ (because the units are calculated based on per units). Also, in sensitive cases, this value can be considered a hyperparameter, and trial and error can obtain a relatively good value. In this paper, a random value in the range is used. The trained LSTM has a lookback $p = 20$, a lookahead
$ = 1$. Two consecutive hidden recurrent layers are fully
connected, with 32 and 64 LSTM units, respectively, a dense
output layer with three neurons. We train the prediction model with mean squared error (MSE) loss, Adam optimizer using a learning rate of 0.0001, and a batch size of 32. The training was performed for 50 epochs with early stopping.

\subsection{Fault Classification}
In data mining and statistics, the KNN classification method is a case-based learning algorithm that depends on a distance or similarity function, such as the Euclidean distance function; because of its simple implementation and significant classification performance, it is a popular method. By the use of predicted voltage ${{V^*}_{inv}}_{a,b,c}(t+1)$, the goal of this part in the control scheme is 1) finding out whether fault happens or not, 2) if it is a fault, which type of fault is.
${{V^*}_{inv}}_{a,b,c}(t+1)$ will be classified into one of 11 classes of faults or normal class. These fault classes are as follows: single-phase A to ground (a-g), single-phase B to ground (b-g), single-phase C to ground (c-g), two-phase A and B (a-b), two-phase B and C (b-c), two-phase C and A (c-a), two-phase A and B to ground (a-b-g), two-phase B and C to ground (b-c-g), two-phase C and A to ground (c-a-g), three-phase A, B and C (a-b-c), three-phase A, B and C to ground (a-b-c-g), and normal. Then, according to the determined label, the input voltage of the PWM unit will be specified. Thus, if this classifier detects the voltage as one of the faults, the next part of the control scheme will be activated, which is responsible for correcting this signal. Otherwise, the controller should send the predicted voltage into PWM unit. The confusion matrix presented in Fig. 5  is the performance measurement for the KNN classification model.  In this model, $k=5$ is considered, which resulted in an accuracy of more than 99\%. 
\begin{figure}[h]
\centerline{\includegraphics[width=9cm,height=7cm]{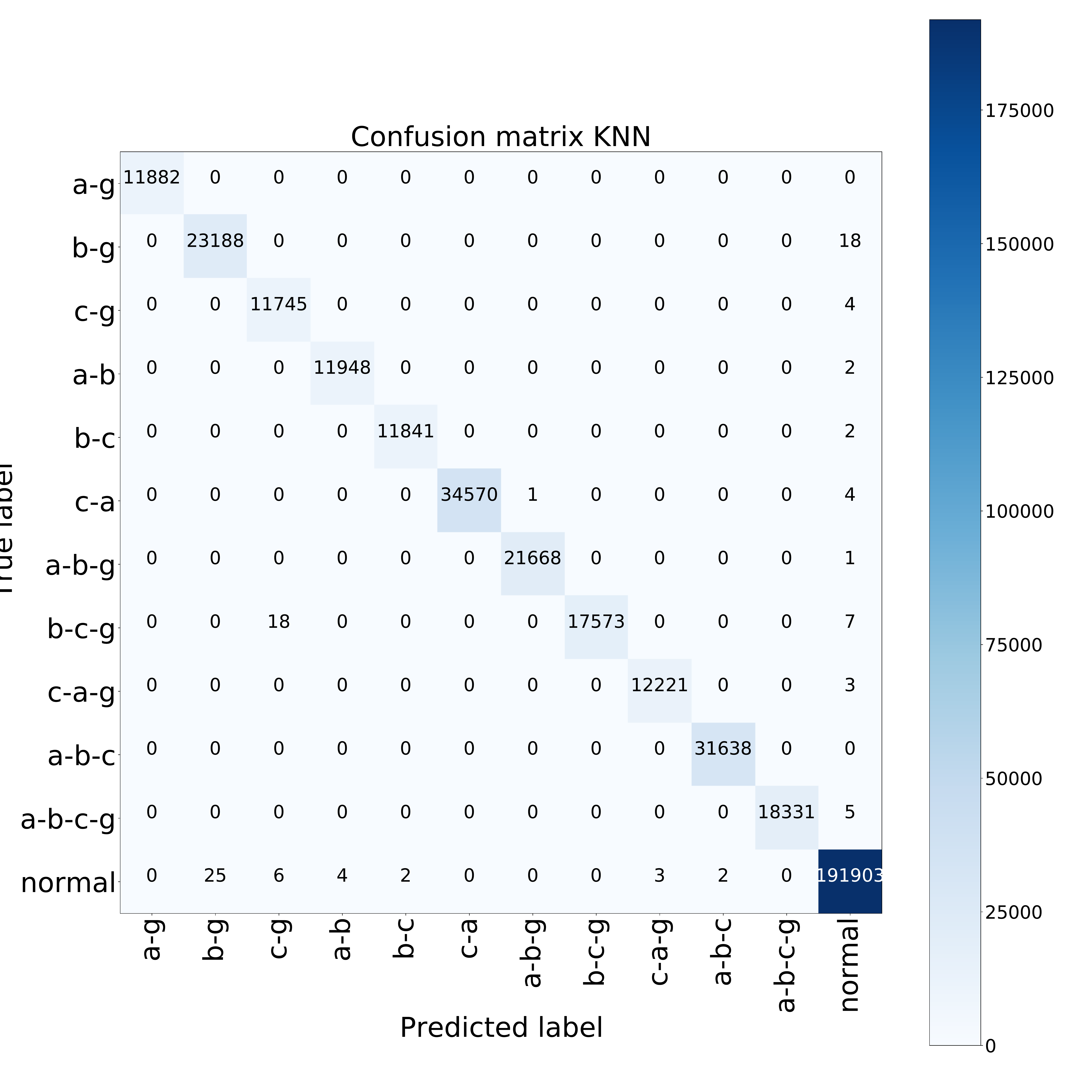}}
\caption{Confusion matrix of KNN classifier.}
\vspace*{-4mm}
\label{fig}
\end{figure}
\subsection{Signal Correction}
If any fault is detected, this part of the control scheme will be activated by a signal from the KNN classifier. The presence of this network in the control scheme ensures that the system will not be disrupted, and maintain its normal operation in the occurrence of a fault. This network which trained only on normal data, is a multi-output regression implemented directly via NN by specifying the number of target variables as the number of output layer nodes. The inputs of this network are the grid voltage $({V_{g}}_{a,b,c})$ and the inverter current $({I_{inv}}_{a,b,c})$, which produces the inverter output voltage $({{V^*}_{inv}}_{a,b,c})$. 
Regarding the network structure, the trained model consists of four layers: an input layer with six neurons, two dense hidden layers with 64 and 128 neurons, and an output layer with three neurons. We train the model with MSE loss, Adam optimizer using a learning rate of 0.001, and a batch size of 16. The training was performed for 30 epochs with early stopping. The ReLU activation function is used in the hidden layers, and the hyperbolic tangent activation function for the output layer; the hyperbolic tangent makes the output signals of this network more accurate and closer to the sine than the output signals of the model-based controller. 
\begin{table}[h]
\caption{Faults duration and parameters for both training and test sets}
\begin{center}

\begin{tabular}{ccccccc}

  \midrule
   Fault & \begin{tabular}{@{}c@{}}Start\\Point  \end{tabular} & \begin{tabular}{@{}c@{}}End\\Point  \end{tabular} &   \small \begin{tabular}{@{}c@{}c@{}}Neutral\\Ground \\ Resistance\end{tabular}   &  \small \begin{tabular}{@{}c@{}c@{}}Phase\\Neutral \\ Resistance\end{tabular} &application \\
  \hline
    a-g      &  0.4 & 0.5 & 0.08 & 0.08&train\\  
  \hline
  b-g      & 0.6 & 0.7& 0.08 & 0.08&train\\  
  \hline
  c-g      & 0.85 & 0.95 & 0.08 & 0.08&train\\ 
    \hline
  a-b     & 1.05 & 1.15 & 0.3 & -&train\\ 
    \hline
  b-c      & 1.4 & 1.6& 0.1 & -&train \\ 
    \hline
  c-a      & 1.7 & 1.85 & 0.6 & -&train \\ 
    \hline
  a-b-g  & 2 & 2.1 & 0.08 & 0.08&train \\ 
    \hline
  b-c-g  & 2.5 & 2.8 & 0.1 & 0.1&train\\ 
    \hline
  c-a-g  & 3 & 3.15 & 0.5 & 0.5&train\\ 
    \hline
  a-b-c  & 3.5 & 3.7  & 0.3 & -&train\\ 
      \hline
  a-b-c-g   & 3.85 & 3.95 & 0.3 &0.3&train \\ 
    \hline
  a-g      & 1 & 1.22 & 0.3 & 0.3&test\\  
  \hline
  b-g      & 2.62 & 2.75  & 0.2 & 0.2&test\\  
  \hline
  c-g      & 0.32 & 0.5  & 0.1 & 0.1&test\\ 
    \hline
  a-b      & 1.12 & 1.19     & 0.07 & -&test\\ 
    \hline
  b-c      & 2.22 & 2.3     & 0.3 & -&test \\ 
    \hline
  c-a      &1.89 & 2.17     & 0.1 & -&test \\ 
    \hline
  a-b-g      & 0.1 & 0.18     & 0.25 & 0.25&test \\ 
    \hline
  b-c-g      & 2.41 & 2.5     & 0.3 & 0.3&test\\ 
    \hline
  c-a-g     & 3.7 & 3.82     & 0.08 & 0.08&test\\ 
    \hline
  a-b-c      & 1.7 & 1.81     & 0.1 & -&test\\ 
      \hline
  a-b-c-g      & 0.72 & 0.8     & 0.5 & 0.5 &test\\
  \bottomrule
\end{tabular}
\end{center}
\end{table}
\vspace*{-5mm}
\begin{figure}[h]
 \centering
     \includegraphics[width=0.5\textwidth,height =185mm ]{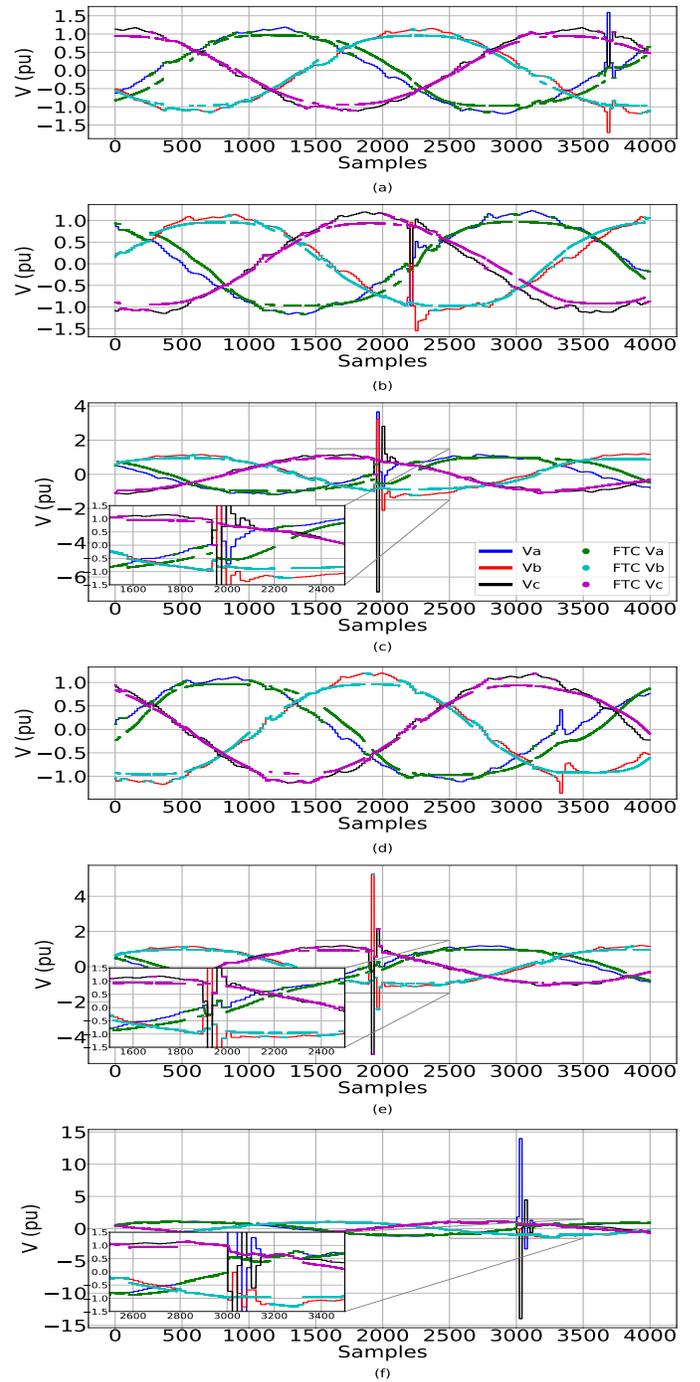}
     \caption{Inverter voltage during single-phase to ground and two-phase faults ($V_a$, $V_b$, $V_c$) and proposed control scheme output ($FTC$ $V_a$, $FTC$ $V_b$, $FTC$ $V_c$).}
    %  \label{fig:y equals x}
 \end{figure}
%  \vspace*{-5mm}
 \begin{figure}[h]
     \centering
     \includegraphics[width=0.5\textwidth,height =175mm]{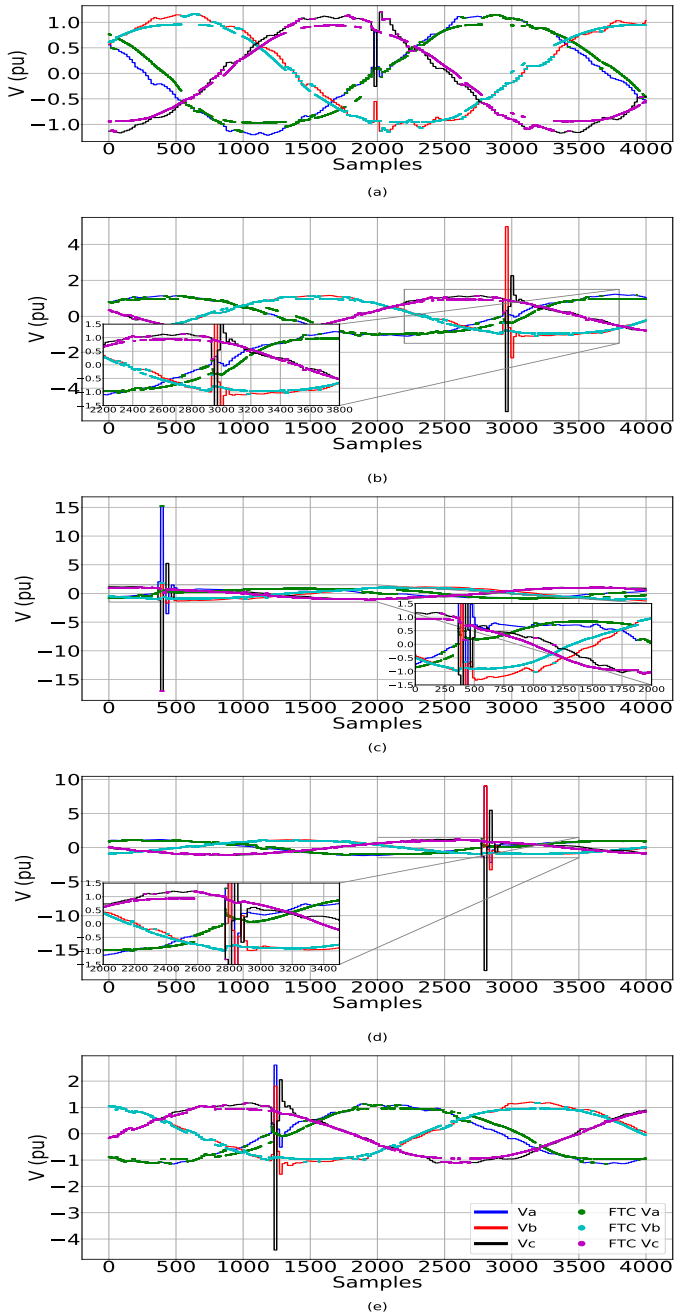}
     \caption{Inverter voltage during two-phase to ground and three-phase faults ($V_a$, $V_b$, $V_c$) and proposed control scheme output ($FTC$ $V_a$, $FTC$ $V_b$, $FTC$ $V_c$).}
    %  \label{fig:three sin x}
    %  \vspace*{-5mm}
 \end{figure}

%  \begin{figure}[h]
%      \centering
%      \includegraphics[width=88mm,height =140mm]{2phase2g.pdf}
%      \caption{Inverter voltage during two-phase to ground faults and proposed control scheme output.}
%     %  \label{fig:five over x}
%  \end{figure}
% \begin{figure}[h]
%  \centering
%      \includegraphics[width=88mm,height =90mm ]{3phase.pdf}
%      \caption{Inverter voltage during three-phase and three-phase to ground faults and proposed control scheme output.}
%     %  \label{fig:y equals x}
%  \end{figure}
\section{Simulation Results}

In order to evaluate the proposed method performance for FDD and FTC in grid-tied DER inverters, first, an inverter and a conventional controller based on [9] were simulated in MATLAB/Simulink. The inverter parameters for the output filter are $R_1 = R_2 = 0.5$ $\Omega$, $C = 4.5$ $\mu F$, $L_1 = L_2 = 0.09$ $mH$, the DC-link voltage is $500$ $V$, and the switching frequency is $10$ $kHz$. Grid frequency and Grid voltage (line to line rms) are $60$ $Hz$ and $220$ $V$ respectively. Second, all grid faults based on Table \RomanNumeralCaps{1} were applied, and training and test set were collected. 

Next, using python libraries such as TensorFlow, Keras, Scikit-Learn, etc., the networks used in this control structure are first trained and validated through the training set and then tested on utterly new data from scenarios 1 and 2. In the training phase, the K-fold cross-validation method is used to prepare training and validation data. In this method, the training data set is randomly divided into $k =10$ subsets. Then, at each stage, $k-1$ of these subsets is considered as a training dataset and one as a validation dataset.

Case studies are carried out for two scenarios; in the first one, compared with the training set, the duration and arrangement of faults have changed. In scenario 2, in addition to changes in the duration and arrangement, the values of the phase-natural resistor and natural-ground resistor have also changed. Due to the massive amount of sample data, the average results of mean absolute error (MAE) for all types of fault are demonstrated in Table \RomanNumeralCaps{2}. MAEs in this table show that the proposed control scheme accurately detects all grid faults, especially faults with significant changes in amplitude.

Moreover, samples for a limited period during fault occurrence (4000 samples) are selected and displayed in Figs. 6 and 7. The proposed controller output in these figures verified its abilities to prevent inverter challenging grid faults and system failure. These figures show that using the proposed FDD and FTC scheme can guarantee the inverter not suffering from any grid faults and failure. Considering the LSTM network, this control scheme can ensure that the system will not be disrupted and maintain its normal operation even if a fault occurs abruptly or continuously over a limited period. 
\begin{table}[h]
\caption{MAEs for different scenarios and different faults}
\begin{center}

\begin{tabular}{ccccccc}
\hline
\textbf{Fault}&\multicolumn{3}{c}{\textbf{Scenario 1}} &\multicolumn{3}{c}{\textbf{Scenario 2}} \\
\cline{2-7} 
\textbf{Type} & \textbf{\textit{$V_a$}}& \textbf{\textit{$V_b$}}& \textbf{\textit{$V_c$}} &\textbf{\textit{$V_a$}}& \textbf{\textit{$V_b$}}& \textbf{\textit{$V_c$}}\\
\hline
a-g& 0.121 &	0.069 &	0.067 &	0.141 &	0.062 &	0.83 \\
\hline
b-g& 0.111 &	0.811 &	0.145 &	0.11 &	0.075 &	0.107 \\
\hline
c-g& 0.234&	0.128&	0.084&	0.24&	0.191&	0.103 \\
\hline
a-b& 0.142&	0.072&	0.058&	0.19&	0.09&	0.081 \\
\hline
b-c& 0.159&	0.137&	0.105&	0.149&	0.102&	0.099 \\
\hline
c-a& 0.122&	0.068&	0.064&	0.141&	0.075&	0.075 \\
\hline
a-b-g& 0.193&	0.161&	0.136&	0.157&	0.082&	0.086 \\
\hline
b-c-g& 0.124&	0.092&	0.069&	0.11&	0.074&	0.06 \\
\hline
c-a-g& 0.198&	0.073&	0.083&	0.404&	0.17&	0.265 \\
\hline
a-b-c& 0.134&	0.067&	0.89&	0.149&	0.088&	0.105 \\
\hline
a-b-c-g& 0.181&	0.089&	0.103&	0.156&	0.074&	0.085 \\
\hline
% \multicolumn{4}{l}{$^{\mathrm{a}}$Sample of a Table footnote.}
\end{tabular}
\label{tab1}
\end{center}
\end{table}

\section{Conclusion}
This paper proposes an integrated FDD and FTC strategy for inverter interfaced DER by using supervised ML methods. We predict the next step of inverter input voltage by using LSTM networks and then applying the KNN classifier, the situation of a fault occurring or normal voltage is recognized. Furthermore, the type of fault will be determined. In case of any grid fault, an NN-based multi-output regression generates a normal signal. Therefore, the inverter never faces system failures. The simulations presented by this paper affirm the proposed advantages.
% Future work will investigate: The proposed control scheme has satisfying accuracy in predicting, detecting, and correcting grid faults, and besides, the simulation results affirm it. However, this control scheme does not have the ability to detect whether the inverter is disconnected or connected to the grid because the multi-output regression learning network is trained only with the normal data of inverters' operation mode.

\end{document}